\documentstyle[11pt]{article}
\newcommand{\dd}{{\rm d}}
\def\dimq{{^{[q]}}}
\def\dimqm1{{^{[q-1]}}}
\def\dimp{{^{[p]}}}
\def\dimpm1{{^{[p-1]}}}
\def\dimqmp{{^{[q,p]}}}
\def\dimqmpm1{{^{[q,p-1]}}}
\def\dimq{{^{[q]}}}

\def\Oqm1{{\Omega^{^{[q-1]}}}}

\def\Opm1{{\Omega^{^{[p-1]}}}}
\def\Oqmp{{\Omega^{^{[q-p]}}}}
\def\Oqmpm1{{\Omega^{^{[q-p-1]}}}}
\def\thickness{\epsilon}
\def\cutoff{\Delta}
\def\smooth{{_{\lbrace\thickness\rbrace}}}

\def\calG{{\cal G}}

\def\profil{{\cal D}}
\def\greenreg{{\cal G}}
\def\greenf{G}
\def\greenregperp{\,{}^{^\perp}\!{\cal G}}
\def\profilperp{\,{}^{^\perp}\!{\cal D}}
\def\greenfperp{\,{}^{^\perp}\!{G}}

\def\petit{\alpha}
\newcommand{\gta}{\mathrel{%
   \rlap{\raise 0.511ex \hbox{$>$}}{\lower 0.511ex \hbox{$\sim$}}}}
\newcommand{\lta}{\mathrel{
   \rlap{\raise 0.511ex \hbox{$<$}}{\lower 0.511ex \hbox{$\sim$}}}}

\def\calH{{\cal H}}
\def\calI{{\cal I}}
\def\calJ{{\cal J}}
\begin{document}

\title{\bf GRADIENT FORMULA FOR LINEARLY SELF-INTERACTING BRANES}

\author{Brandon Carter$^1$, Richard A. Battye$^{2,3}$ and 
Jean--Philippe Uzan$^4$ \\ \\ 
$^1$ LUTH, B\^at. 18, Observatoire de Paris, F-92195 Meudon, France. \\  \\
$^2$ Jodrell Bank Observatory, Macclesfield, Cheshire SK11 9DL, U.K. \\  \\ 
$^3$ Department of Physics and Astronomy, University of Manchester, \\ 
Schuster Laboratory,  Brunswick St, Manchester M13 9PL, U.K. \\ \\ 
$^4$ Laboratoire de Physique Th\'eorique, CNRS--UMR~8627, B\^at. 210, \\ 
Universit\'e Paris XI, F-91405 Orsay Cedex, France.}

\date{3rd April 2002}

\maketitle

\begin{abstract}
The computation of long range linear self-interaction forces in string 
and higher dimensional brane models requires the evaluation of the 
gradients of regularised values of divergent self-interaction potentials. 
It is shown that the appropriately regularised gradient in directions 
orthogonal to the brane surface will always be obtainable simply by 
multiplying the regularised potential components by just half the trace 
of the second fundamental tensor, except in the hypermembrane case for 
which the method fails. Whatever the dimension of the background this
result is valid provided the codimension is two (the hyperstring case)
or more, so it can be used for investigating brane-world scenarios with 
more than one extra space dimension.
\end{abstract}

\section{Introduction}\label{sec_intro}

In recent studies of the linear self-interaction of
superconducting~\cite{carter97} and other~\cite{carter98,carter00}
cosmic string models in a standard (3+1) dimensional background
spacetime,  it has been found that the relevant  divergences can be
conveniently dealt with by the application of a
simple universal {\it
gradient  formula}. This formula can be used to obtain the effective
self-force from the  appropriately regularized self-field and the
precise form of this gradient formula does not depend on the physical
nature of the fields to which the string may be coupled. These could,
for example, be of the ordinary (observationally  familiar)
electromagnetic~\cite{carter97}, and linearised
gravitational~\cite{carter98} kinds, but might also be of the
theoretically predicted dilatonic and axionic kinds~\cite{carter00}.

The purpose of this work is to generalise this gradient formula to higher 
dimensional branes in higher dimensional backgrounds. A particular 
motivation for doing this is to generalise the study of the original kind 
of brane-world scenario~\cite{BW} not just by dropping the postulate of 
reflection symmetry~\cite{Davis01,CU01}, but by allowing for more than one 
extra dimension~\cite{high,high2,high3}, so as to obtain models 
consisting of a 3-brane embedded in a spacetime of  dimension greater 
than 5. More generally, our results are applicable to any $p$-brane in a
 $(q+1)$ dimensional background spacetime, so long as the codimension 
$(q-p)$ is two (the ``hyperstring'' case, which includes ordinary strings 
in a (3+1) dimensional spacetime) or more. The method given here does 
however fail (due to long range ``infrared'' divergences) when the 
co-dimension is only one (the ``hypermembrane'' case, which includes 
ordinary membranes in a (3+1) dimensional spacetime), but in that case it 
is not really needed, since the short range ``ultraviolet'' divergences 
will have a relatively innocuous form that can be dealt with in terms of 
simple jump discontinuities.

The reason why one needs a gradient formula is that the relevant forces 
are typically obtained as gradients of potentials. Outside the brane
worldsheet these potentials will be well behaved fields, but in the 
thin brane limit they will be singular on the worldsheet itself.
One therefore needs to regularise these fields by some appropriate 
ultra-violet cut-off procedure, typically involving a length scale, 
$\thickness$ say, that can be interpreted as characterising the 
underlying microstructure. When one has obtained the appropriately 
regularised potential on the worldsheet (which represents the 
cross-sectional average over an underlying microstructure) the next 
problem is to obtain the corresponding force by evaluating its gradient. 
The gradient components in directions tangential to the worldsheet 
can be obtained directly just by differentiating the regularised 
(macroscopic averaged) potential. The trouble is that, since the 
support of the regularised potential is confined to the worldsheet, 
differentiation in orthogonal directions is not directly meaningful.

To obtain the necessary orthogonal gradient components it is therefore
necessary, in principle, to go back to the underlying microstructure
and perform the ultra-violet limit process again, a rather arduous task
that has frequently been performed independently by different authors
in a wide range different physical contexts.

The present work has been prompted by the observation~\cite{carter97}
that, in the case of a string in an ordinary 3-dimensional background, 
the effect of the orthogonal gradient operation will always turns out 
to be equivalent just to multiplication of the regularised value of
the relevant potential field  by {\it exactly} half the extrinsic 
curvature vector $K^\rho$, which  is defined to be the trace
\begin{equation}\label{19}
K^\rho=g^{\mu\nu}{K_{\mu\nu}}^{\!\rho}\, ,
\end{equation}
of the second fundamental tensor ${K_{\mu\nu}}^\rho$ of the worldsheet. 
The recognition of this simple universal rule makes it unnecessary to go 
back and perform the calculation over again every time such a problem 
arises in a different physical context. On the basis of dimensional 
considerations, and of the need to respect local Lorentz invariance, one 
would expect that multiplication by a factor proportional to $K^\rho$ 
would inevitably be what is required. But what is not so obvious in 
advance is whether the relevant proportionality factor should still 
always be {\it exactly a half} even for branes and backgrounds of higher 
dimensions.

The present work addresses this question and provides an unreservedly 
affirmative reply whenever $0< p < q-1$, on the basis of the consideration 
that whatever the relevant  factor may be for a generic evolution of a brane 
of given space dimension $p$ in a back ground of given space (as opposed to 
spacetime) dimension $q$, this factor must evidently remain the same for  
any {\it static} configuration of the brane. 

By restricting our attention to static configurations, for which a rigourous
analysis is technically much  simpler than in the dynamic case, what we
have succeeded in showing here is that that  -- provided the codimension 
$(q-p)$ is two (the ``hyperstring'' case) or more, so that the quantities
involved are at worst logarithmically divergent in the ``infra red''
(which excludes the strongly divergent case  of a membrane, but admits
the case of a string in 3 dimensions) -- the appropriate factor is
indeed {\it always exactly a half}, no matter how high the dimension of the
brane  or the background. However our analysis also shows that this easily
memorable result will no longer hold in the strongly ``infra red'' divergent 
extreme case of a ``hypermembrane'' (meaning a brane with codimension 
one, so that its supporting worldsheet is a hypersurface). What the present
approach can not do is to provide any evidence at all, one way or the other, 
about value of the factor in question in the opposite extreme case of a 
zero-brane, meaning a simple point particle.

The article is organised as follows. In Section~\ref{sec_geo}, we
recall the definitions of the basic geometrical quantities, in particular
the second fundamental tensor, that will appear in our
discussion.  We introduce the Poisson equation, both in the thin brane
limit and for regularised configurations in Section~\ref{sec_regul}
and discuss its solution in Section~\ref{sec_sol}. The cross-sectional
average of the field and its gradient have a zero-order contribution,
in the case of a flat brane, that is evaluated in
Section~\ref{sec_flat} and we illustrate our formalism on the choice
of a canonical profile function in Section~\ref{sec_cano}. The
cross-sectional average of the gradient of the field vanishes in the
flat configuration and our goal is to evaluate its value at first
order in the curvature. After defining the procedure of averaging in
Section~\ref{sec_weight}, we study the effect of the bending of the
brane on this procedure in
Section~\ref{sec_curv_avrg}. Section~\ref{sec_potcurv} finalises our
demonstration by computing the cross-sectional average of the gradient
of the field. A summary of the notation and some useful integrals are 
presented in the appendix.

\section{Brane embedding geometry}\label{sec_geo}

Since it will play a central role in our analysis, it is worthwhile
to recall the definition and some basic properties of the second
fundamental tensor ${K_{\mu\nu}}^\rho$.

We consider a $p$-dimensional surface (a $p$-brane) with internal 
coordinates $\sigma^i$ (with $i=1\ldots p$) smoothly embedded in a 
higher $q$-dimensional background endowed with a metric $g_{\mu\nu}$ 
with respect to coordinates $x^\mu$ (with $\mu=1\ldots q$). If the 
location of the surface is given in terms of a set of embedding 
functions $\bar x^\mu(\sigma^i)$, the internal surface metric, also 
referred to as {\it first fundamental tensor}, can be expressed as
\begin{equation}\label{11}
\gamma_{ij}= g_{\mu\nu}{\partial \bar x^\mu\over\partial\sigma^i}
{\partial \bar x^\nu\over\partial\sigma^j}\,.
\end{equation}
Provided that this metric is non-singular (in the sense that the
determinant of the component matrix does not vanish), as will
always be the case for a strictly timelike surface in a Lorentz
signature background spacetime, or for any surface if the
background metric is positive definite, then it will have a well
defined inverse with components $\gamma^{ij}$. It can be used to
raise internal coordinate indices in the same way as the inverse
background metric $g^{\mu\nu}$ is used for raising background
coordinate indices. For any such non-singular embedding, $\gamma^{ij}$ 
can be mapped into the background tensor
\begin{equation}\label{13}
\gamma^{\mu\nu}=\gamma^{ij}{\partial \bar
x^\mu\over\partial\sigma^i} {\partial \bar
x^\nu\over\partial\sigma^j}\, .
\end{equation}
The corresponding mixed tensor $\gamma^\mu_{\,\nu}$ acts on vectors as
the natural (rank $p$) surface tangential projector operator, while
its (rank $q-p$) complement
\begin{equation}\label{14}
\perp^{\!\mu}_\nu=g^\mu_{\,\nu}-\gamma^\mu_{\,\nu}\, ,
\end{equation}
acts similarly as the corresponding orthogonal projection operator.

Fields with support confined to the $p$-surface ${\cal S}^{(p)}$, such 
as $\gamma^\mu_{\,\nu}$ and $\perp^{\!\mu}_\nu$, can not be directly 
subjected to the unrestricted operation of partial differentiation with 
respect to the background coordinates, but only to differentiation in 
tangential directions, as performed by the operator
\begin{equation}\label{00}
\overline\nabla_{\!\mu}=\gamma^\rho_{\,\mu}\nabla_{\!\rho} \, ,
\end{equation}
where $\nabla_\rho$ is the usual covariant derivative associated with
$g_{\mu\nu}$. The action of this differential operator on the first 
fundamental tensor defines the {\it second fundamental tensor} of the 
$p$-surface, according to the specification~\cite{carter95}
\begin{equation}\label{16}
{K_{\mu\nu}}^{\!\rho}=\eta_{\nu\sigma}\nabla_{\!\mu}
\eta^{\rho\sigma} \, .
\end{equation}
As a non-trivial integrability condition, it satisfies the generalised 
Weingarten symmetry condition
\begin{equation}\label{16a}
{K_{\mu\nu}}^{\!\rho}={K_{\nu\mu}}^{\!\rho}\, .
\end{equation}
It is also $p$-surface orthogonal on its last index and tangential on 
the other two, that is,
\begin{equation}\label{17}
{K_{\mu\nu}}^{\!\sigma}\eta^\rho_{\,\sigma}=0=
\perp^{\!\sigma}_{\,\mu}{K_{\sigma\nu}}^{\!\rho}\, ,
\end{equation}
so that it has only a single non-identically vanishing self-contraction, 
namely the extrinsic curvature vector introduced in equation (\ref{19}).

\section{Regularisation of brane supported source 
distribution}\label{sec_regul}

Since we are essentially concerned with the treatment of ``ultraviolet'' 
regularisation, the effects of long range background curvature will be 
unimportant and can thus be neglected. As the framework for our analysis, 
it will be sufficient to consider a $q$-dimensional background space 
with a flat, static, positive definite metric $g_{\mu\nu}$. It is
therefore possible to choose a linear (but not necessarily orthogonal) 
system of space coordinates $x^\mu$ so that the background metric will 
be constant, that is, $\partial g_{\mu\nu}/\partial x^\rho=0$.

This choice has different implications. First,  the tangentially
projected covariant derivative (\ref{00}) will take the simple form
\begin{equation}\label{15}
\overline\nabla_{\!\mu}=\gamma^\rho_{\,\mu} {\partial\over\partial
x^\rho}=g_{\mu\nu}\gamma^{ij} {\partial \bar
x^\nu\over\partial\sigma^i} {\partial\over\partial \sigma^j}\, .
\end{equation}
Second, in the linear system (for, example Maxwellian or linearised
gravitational), with which we are concerned here, each tensorial 
component decouples and evolves independently like a simple scalar 
field. Thus, it will be sufficient for our purpose to concentrate on 
the prototype problem of finding a scalar field, let us say $\phi\{x\}$ 
(introducing the systematic use of curly brackets to indicate functional 
dependence), that satisfies the corresponding generalised Poisson 
equation to which the linear wave equation will be reduced in the static 
case to be dealt with here. Third, the background spacetime is assumed to 
be static so that the generalised Laplacian $\nabla^\mu\nabla_{\!\mu}$ is 
to be regarded as as the static projection of what would be a Dalembertian 
type wave operator in a more complete $(p+1)$-dimensional spacetime
description.

\subsection{Generalised Poisson equation and its
 solutions in the thin brane limit}

In unrationalised units for, let us say, a repulsive scalar source
distribution $\rho\{x\}$  the generalised Poisson 
equation takes the form
\begin{equation}\label{1}
\nabla^\mu\nabla_{\!\mu}\,\phi = -\Oqm1\rho\{x\} \, ,
\end{equation}
where the Laplacian  operator is simply given by
\begin{equation}\label{0}
\nabla^\mu\nabla_{\!\mu}=g^{\mu\nu}{\partial^2\over\partial
x^\mu\partial x^\nu}\, ,
\end{equation}
where $\Oqm1$ is the surface area of a unit $(q-1)$-sphere in
the $q$-dimensional background. The latter is given by the 
well known formula
\begin{equation}\label{2}
\Oqm1=\frac{2\pi^{q/2}}{\Gamma\{{q/2}\}}
\,,
\end{equation}
in which $\Gamma\{z\}$ is the usual Eulerian Gamma function 
(as specified in the appendix) which satisfies the recursion 
relation $\Gamma\{n+1\}=n\Gamma\{n\}$ and has the
particular values $\Gamma\{1\}=1$ and $\Gamma\{1/2\}=\sqrt{\pi}$. 
Thus in particular we shall have $\Omega^{[0]}=2$, $\Omega^{[1]}=2\pi$, 
$\Omega^{[2]}=4\pi$, $\Omega^{[3]}=2\pi^2$ and $\Omega^{[4]}=8\pi^2/3$.

The motivation for working with a $p$-brane model of a physical system
is of course to be able to use an economical description in which as 
many as possible of the fields involved are specified just on the 
supporting surface. Evidently, this requires less information than 
working with fields specified over the higher dimensional background. 
However, when long range interactions are involved then, as a price for 
such an economy, divergences are to be expected if the source fields 
are considered to be strictly confined to the supporting $p$-surface. 
To be more specific, the brane supported source distribution 
$\bar\rho\{\sigma\}$ (using a bar as a reminder to indicate quantities 
that are undefined off the relevant supporting surface) will correspond 
to the  background source distribution
\begin{equation}\label{01}
\rho\{x\}=\int\delta^\dimq\{x,\bar x\{\sigma\}\}\,
\bar\rho\{\sigma\} \, \dd{\cal S}^\dimp\, ,
\end{equation}
where $\delta^\dimq\{x,\bar x\{\sigma\}\}$ is the $q$-dimensional 
bi-scalar Dirac distribution. This is characterised by the condition that 
it vanishes wherever the evaluation points $x^\mu$ and 
$\bar x^\mu\{\sigma\}$ are different while nevertheless satisfying the 
unit normalisation condition
\begin{equation}\label{02}
\int \delta^\dimq\{x,\bar x\}\, \dd{\cal S}^\dimq=1 \, .
\end{equation}
We use the notations  $\dd{\cal S}^\dimp$ and  $\dd{\cal S}^\dimq$
respectively for the $p$-dimensional brane surface measure and
$q$-dimensional background space measure. The latter is given by an 
expression of the form
\begin{equation}\label{03}
\dd{\cal S}^\dimq=|g|^{1/2} \dd^\dimq x\, ,
\end{equation}
where the $q$-dimensional volume measure factor $|g|^{1/2}$ is just the 
square root of the determinant of the (positive definite) matrix of 
metric components $g_{\mu\nu}$. Similarly, we have
\begin{equation}\label{12}
\dd{\cal S}^\dimp=|\gamma|^{1/2} \dd^\dimp\sigma\, ,
\end{equation}
where the $p$-dimensional volume measure factor $|\gamma|^{1/2}$ is the 
square root of the determinant of the matrix of surface metric 
components, as given by (\ref{11}). The normalisation condition 
(\ref{02}) is such that the measure factor means that the Dirac 
distributions specified in this way will transform as an ordinary 
bi-scalar, a property that distinguishes it from the corresponding 
Dirac type that would be given by the commonly used alternative 
convention in which the measure factor is omitted. This specifies 
distributions that behave not as ordinary bi-scalars but as weight 
one bi-scalar densities. Of course it does not matter which of these 
conventions is used if, as will be done below, the coordinates are 
restricted to be of the standard orthonormal type with respect to which 
the metric just has the Cartesian unit diagonal form.

For the source distribution given in the thin brane limit by (\ref{01}), 
the corresponding solution of the linear field equation (\ref{1}) is 
given in terms of a Green function by the expression
\begin{equation}\label{23}
\phi\{x\}=\int\greenf^\dimq\{x,\bar x\{\sigma\}\}\,\bar\rho\{\sigma\} \,
\dd{\cal S}^\dimp\, ,
\end{equation}
while the corresponding scalar field gradient -- which is what
determines the resulting force density -- will be given by
\begin{equation}\label{24}
{\partial\phi\over\partial x^\mu}\{x\}=
\int{\partial\greenf^\dimq\over\partial x^\mu}\{x,\bar
x\{\sigma\}\}\,\bar\rho\{\sigma\} \, \dd{\cal S}^\dimp\, .
\end{equation}
The Green function $\greenf^\dimq\{x,\bar x\{\sigma\}\}$ satisfies
the Laplace equation
\begin{equation}\label{24a}
\nabla_\mu\nabla^\mu\greenf^\dimq\{x,\bar x\{\sigma\}\}
=-\Oqm1 \delta^\dimq\{x,\bar x\{\sigma\}\}\,
\end{equation}
subject to the appropriate boundary conditions, which in the standard 
case for an isolated system will simply consist of the requirement that 
the field should tend to zero at large distances from the source. Like 
the Dirac ``function'', it is a bi-scalar (not a bi-scalar density) that
is well behaved as a distribution, but singular as a function wherever
the two points on which it depends are coincident.  Note that the Green
function is determined up to an harmonic function, that is a solution of 
the homogeneous Poisson equation
$\nabla_\mu\nabla^\mu\greenf^\dimq\{x,\bar x\{\sigma\}\}=0$, with the 
appropriate boundary conditions. In particular $\greenf^\dimq$ is
determined up to an additive constant that induces a global shift in the
potential and is not relevant for the computation of its gradient.

\subsection{From the thin brane to a regular thick brane}

Although convenient for many formal calculational purposes, the drawback 
to the use of the strict thin brane limit as represented in terms of such 
singular Dirac and Green distributions is that the ensuing value of the 
corresponding field $\phi$ and its gradient $\partial\phi/\partial x^\mu$ 
will also be singular on the brane $p$-surface, which unfortunately is 
just where one needs to evaluate them for purposes such as computing the 
corresponding energy and force density.

The obvious way to get round this difficulty -- without paying the cost 
of introducing extra degrees of freedom -- is to replace (\ref{01}) by an 
expression of the form
\begin{equation}\label{04}
\widetilde\rho_\smooth\{x\}=\int\profil^\dimq_\smooth\{x,\bar
x\{\sigma\}\}\,\bar\rho\{\sigma\} \, \dd{\cal S}^\dimp\, ,
\end{equation}
in which the strict Dirac distribution $\delta^\dimq\{x,\bar x\}$ is
replaced by a regular bi-scalar profile function,
$\profil^\dimq_\smooth\{x,\bar x\}$ say, characterised by some
sufficiently small but finite smoothing length scale, $\thickness$ say,
which is subject to the normalisation condition
\begin{equation}\label{05}
\int\profil^\dimq_\smooth\{x,\bar x\}\, \dd{\cal S}^\dimq =1 \, .
\end{equation}
In order to respect the geometric isotropy and homogeneity of the
background space, this regular ``profile function''
$\profil^\dimq_\smooth\{x,\bar x\}$ can depend only on the relative
distance, $s$ say, that is defined -- in the flat background under
consideration -- by
\begin{equation}\label{6}
s^2= g_{\mu\nu} (x^\mu-\bar x^\mu) (x^\nu-\bar x^\nu)\, ,
\end{equation}
according to some ansatz of the form
\begin{equation}\label{06}
\profil^\dimq_\smooth\{x,\bar x\}=\profil^\dimq_\smooth\{s^2\} \, .
\end{equation}
The functionally singular Dirac distribution can be considered as the
limit of such a function as the smoothing lengthscale tends to zero. 
Such a description corresponds to what may be described as ``fuzzy'' 
brane model characterised by an effective thickness whose order of 
magnitude will be a function of the value of the regularisation parameter 
$\thickness$. For some kinds of physical application one might wish to 
adjust the regularisation ansatz so to make it agree as well as possible 
with the actual internal structure (if known) of the extended physical 
system under consideration. The spirit of the present work is, however, 
rather to treat the regularisation parameter just as a provisional freely 
variable quantity for use at an intermediate stage of an analysis process
whereby it would ultimately be allowed to tend to zero so as to provide a 
strict ``thin'' brane limit description. In order for our results to be 
meaningful for this latter purpose, it is important that they should be 
insensitive to the particular choice of the smoothing ansatz that is 
postulated. We shall take care to verify this

\section{Field solution for regularised source distribution}
\label{sec_sol}

For any source distribution of the regularised form (\ref{04}), the 
function, $\greenreg^\dimq_\smooth\{s^2\}$ that is the solution (with 
the appropriate boundary conditions) of the corresponding generalisation 
\begin{equation}\label{07}
\nabla^\mu\nabla_{\!\mu}\greenreg^\dimq_\smooth\{s^2\}=
-\Oqm1\profil_\smooth^\dimq\{s^2\} \, ,
\end{equation}
of (\ref{24a}) will act as a generalised Green function. The solution of 
the Poisson equation (\ref{1}) for the generic smoothed source 
distribution (\ref{04}) will thus be given, using the natural 
abbreviation
\begin{equation}\label{08}
s\{x,\sigma\}=s\{x,\bar x\{\sigma\}\}\, ,
\end{equation}
 by
\begin{equation}\label{09}
\widetilde\phi_\smooth\{x\}=\int\greenreg^\dimq_\smooth\{s\{x,\sigma\}\}
\,\bar\rho(\sigma)\, \dd{\cal S}^\dimp\, .
\end{equation}

The gradient of this scalar field is evidently given by
\begin{equation}\label{010}
{\partial\widetilde\phi_\smooth\over\partial x^\mu}\{x\}=\int{\partial
\greenreg_\smooth^\dimq\over\partial x^\mu}
\{s^2\{x,\sigma\}\}\,\bar\rho\{\sigma\}\, \dd{\cal S}^\dimp\, ,
\end{equation}
in which the integrand can be evaluated using the formula
\begin{equation}\label{012}
{\partial\greenreg_\smooth^\dimq\over\partial x^\mu}=2
s_\mu{\dd\greenreg_\smooth^\dimq\over \dd s^2}\, ,
\end{equation}
where $s^\mu=x^\mu -\bar x^\mu$.  One can evidently go on in the same
way to obtain the second derivative
\begin{equation}\label{013}
{\partial^2\widetilde\phi_\smooth\over\partial x^\mu\partial x^\nu}\{x\}
=\int{\partial^2 \greenreg_\smooth^\dimq\over\partial x^\mu\partial x^\nu}
\{s^2\{x,\sigma\}\}\,\bar\rho\{\sigma\}\, \dd{\cal S}^\dimp\, .
\end{equation}
in which the integrand can be evaluated using the formula
\begin{equation}\label{014}
{\partial^2\greenreg_\smooth^\dimq\over\partial x^\mu\partial x^\nu}
= 2g_{\mu\nu}{\dd\greenreg_\smooth^\dimq\over \dd s^2}
+4s_\mu s_\nu {\dd^2\greenreg_\smooth^\dimq\over (\dd s^2)^2}\, .
\end{equation}
It is to be noted that, by contraction of (\ref{014}), the Laplace
equation (\ref{07}) for $\greenreg_\smooth^\dimq$ takes the form
\begin{equation}\label{015}
{\dd\over \dd s^2}\left(s^q{\dd\greenreg_\smooth^\dimq\over\dd s^2}
\right)= -{\Oqm1\over 4}s^{q-2}\profil_\smooth^\dimq\{s^2\} \, .
\end{equation}

If one introduces the monotonically increasing function 
${\calI}_\smooth^\dimq\{s^2\}$ by
\begin{equation} \label{017}
{\calI}_\smooth^\dimq\{s^2\}\equiv{\Oqm1}\int_0^{s}
u^{q-1} \profil_\smooth^\dimq\{u^2\}\, \dd u\, ,
\end{equation}
then it will satisfy
\begin{equation} \label{017a}
{\dd{\calI}_\smooth^\dimq\over \dd s^2}=
{\Oqm1\over 2}s^{q-2}\profil_\smooth^\dimq\{s^2\}
\, .\end{equation}
By substituting into (\ref{015}), it follows that we shall have
\begin{equation}\label{016}
{\dd\greenreg_\smooth^\dimq\over \dd s^2}=-{{\calI}_\smooth^\dimq\{s^2\}
\over 2s^q}\,,
\end{equation}
and hence
\begin{equation} \label{016a}
{\dd^2\greenreg_\smooth^\dimq\over (\dd s^2)^2}={q\,
{\cal I}_\smooth^\dimq\{s^2\}
\over 4s^{q+2}}-{\Oqm1\profil_\smooth^\dimq\{s^2\}\over 4 s^2}
\, .
\end{equation}
The normalisation condition (\ref{05}) implies by (\ref{017}) that the
function ${\calI}_\smooth^\dimq\{s^2\}$ should satisfy the condition
\begin{equation}\label{018}
{\calI}_\smooth^\dimq\{s^2\}\rightarrow 1
\end{equation}
in the limit $s^2 \rightarrow \infty$, for all $\thickness$. It
follows that any admissible ansatz for the smoothing profile will
entail an asymptotic behaviour characterised by
\begin{equation}\label{019}
{\dd\greenreg_\smooth^\dimq\over \dd s^2}\sim -{1\over 2s^q}
\end{equation}
as $s^2\rightarrow\infty$. The corresponding asymptotic formula for the
regularised Green function $\greenreg_\smooth^\dimq\{s^2\}$ will therefore 
be given, as long as $q>2$, by
\begin{equation}\label{020}
\greenreg_\smooth^\dimq\{s^2\}\sim {s^{2-q}\over q-2}
\end{equation}
as $s\rightarrow\infty$.  In the particular case $q=2$, the integral over
$s^2$ is marginally convergent, with asymptotic behaviour characterised by
\begin{equation}\label{020b}
\greenreg_\smooth^{^{[2]}}\{s^2\}\sim-\frac{1}{2}\ln \{s^2\}
\end{equation}
as $s^2\rightarrow\infty$.

\section{Evaluation for flat brane configurations}\label{sec_flat}

Before considering the effect of the curvature of the brane, it will be
instructive to apply the preceding formulae to the case of a flat 
$p$-brane configuration supporting a uniform source distribution.

The uniformity condition means that the value of $\bar\rho$ is
independent of the internal coordinates $\sigma^i$, and the flatness
condition means that the embedding mapping $\sigma^i\mapsto
{\bar x}^\mu\{\sigma^i\}$ of the brane $p$-surface will be expressible 
in a system of suitably aligned Cartesian background space coordinates
$\{{\bar x}^\mu\} = \{{\bar z}^i, {\bar r}^a\}$ (with $\mu=1\ldots q$, 
$i=1\ldots p$, $a=p+1\ldots q$) in terms of a corresponding naturally 
induced system of internal coordinates $\sigma^i$ simply by
\begin{equation}\label{31}
{\bar z}^i =\sigma^i\, ,\hskip 1 cm {\bar r}^a=0 \, .
\end{equation}
With  such a coordinate system, the background metric components will be 
given by
\begin{equation}\label{32}
g_{ij}=\gamma_{ij}, \hskip 1 cm
g_{ia}=0\, ,\hskip 1 cm
g_{ab}=\perp_{ab}\, ,
\end{equation}
where $\gamma_{ij}$ and $\perp_{ab}$ are unit matrices of dimension $p$ 
and $(q-p)$ respectively.

The intrinsic uniformity of the configuration ensures that physical 
quantities depend only on the external coordinates $r^a$, and the isotropy 
of the system ensures that scalar quantities can depend only on the 
radial distance $r$ from the brane $p$-surface given by
\begin{equation}\label{33}
r^2=\perp_{ab}r^a r^b \, .
\end{equation}
In particular, the distance $s$ from a point with coordinates $x^\mu$ on
the transverse $(q-p)$-plane through the origin (that is, with 
$\sigma^i=0$) to a generic point with coordinates $\sigma$ on the flat 
$p$-brane locus will be given simply by
\begin{equation}\label{37}
s^2\{x,\sigma\}=r^2+\sigma^2\, ,
\end{equation}
where
\begin{equation} \label{38}
\sigma^2=\gamma_{ij}\sigma^i\sigma^j\, .
\end{equation}
It can be seen that for a given fixed value of the source density
$\bar\rho$ on the brane, the prescription (\ref{04}) will provide a
radial dependence of the form
\begin{equation}\label{041}
\widetilde\rho\{r^2\}=\profilperp_\smooth^\dimqmp\{r^2\}\,\bar\rho\,,
\end{equation}
where the dimensionally reduced radial profile function 
$\profilperp^\dimqmp_\smooth\{r^2\}$ is given in terms of the original 
$q$-dimensional profile function $\profil^\dimq_\smooth\{s^2\}$ by the 
expression
\begin{equation} \label{042}
\profilperp^\dimqmp_\smooth\{r^2\}=\Opm1\int_0^\infty
 \sigma^{p-1}\profil_\smooth^\dimq \{\sigma^2+r^2\}\,\dd\sigma
\, .\end{equation}
This integral will always be convergent as a consequence of the rapid 
asymptotic fall-off condition that must be satisfied by the function 
$\profil^\dimq\{s^2\}$ in order for it to obey the normalisation 
condition (\ref{05}), which implies that 
\begin{equation}\label{042a}
\Oqmpm1\int_0^\infty  r^{q-p-1}
\profilperp_\smooth^\dimqmp\{r^2\}\, \dd r\,     =1\,.
\end{equation}

Since (\ref{019}) implies that the associated regularised Green function 
$\greenreg^\dimq_\smooth\{s^2\}$ will satisfy the comparatively weak 
asymptotic fall off condition (\ref{020}), the field 
$\widetilde\phi_\smooth$ will be given by
\begin{equation} \label{043}
\widetilde\phi_\smooth\{r^2\}=\bar\rho\, 
\greenregperp_\smooth^\dimqmp\{r^2\}\,,
\end{equation}
using (\ref{09}), where the dimensionally reduced radial Green function 
$\greenregperp_\smooth^\dimqmp\{r^2\}$ is defined by
\begin{equation}\label{044}
\greenregperp_\smooth^\dimqmp\{r^2\}\equiv{\Opm1}
\int_0^{\cutoff} \sigma^{p-1} \greenreg_\smooth^\dimq\{\sigma^2+r^2\}
\, \dd\sigma\, ,
\end{equation}
in the limit where $\cutoff\rightarrow\infty$, a limit that
will be convergent only if  $q>p+2$.

In the prototype case~\cite{carter97,carter98,carter00} of an ordinary 
string with $p=1$ in a background of space dimension $q=3$, the latter
is only marginally too low compared with the dimension of the brane surface 
to achieve convergence. In that case, as in  the more general 
{\it hyperstring} case, that is to say whenever $q=p+2$, it is still 
possible to obtain a useful asymptotic formula by taking the 
upper limit $\cutoff$ to be a finite ``infra-red'' cut off length. 
It can be seen from (\ref{020}) that in this marginally divergent 
codimension-2 case the resulting asymptotic behaviour will be given by
\begin{equation} \label{052}
\greenregperp_\smooth^{^{[q,q-2]}}\{r^2\}\sim {\Omega^{^{[q-3]}}\over q-2}
\ln\left\{{\cutoff\over\sqrt{r^2+\thickness^2}}\right\}\,,
\end{equation}
as $\cutoff\rightarrow\infty$. 

One might go on to try obtain an analogous formula for what, as far as
``infra-red'' divergence behaviour is concerned, is the worst possible 
case of all, namely that of a {\it hypermembrane} (or ``wall'') meaning 
the hyper-surface supported case characterised by $q=p+1$. However in 
this codimension-1 case, the divergence will have the linear form
\begin{equation}\label{052a}
\greenregperp_\smooth^{^{[q,q-1]}}\{r^2\}\sim {\Omega^{[q-2]}
\over q-2}\, \Delta\, ,
\end{equation}
which is particularly sensitive to the choice of cut-off $\Delta$ and 
hence is not very useful.

For physical purposes the potential $\widetilde\phi_\smooth$ will 
typically be less important than its (not so highly gauge dependent) 
gradient, which is what will determine the relevant force density. In 
the flat and uniform configuration considered in this section, the 
gradient $\partial\widetilde\phi_\smooth/ \partial x^\mu$ is not liable 
to the infrared divergence problems that beset the undifferentiated field 
when the background dimension $p$ is not sufficiently high compared with 
the brane dimension $q$.  It is evident from the uniformity of the
configuration that the gradient components in directions parallel to the
brane $p$-surface will vanish, that is, we shall have 
$\partial\widetilde\phi_\smooth/\partial z^i=0$, while starting from the 
expression (\ref{010}), it can be concluded that the orthogonal gradient 
components will be given by
\begin{equation}\label{053}
{\partial\widetilde\phi_\smooth\over\partial r^a}\{r^2\}=\bar\rho\,
 {\partial\greenregperp_\smooth^\dimqmp\over \partial r^a}\{r^2\}
\end{equation}
where, by Eq.~(\ref{016}), even in the cases for which the
undifferentiated integral (\ref{044}) would be divergent, the
required derivative is given by a well defined formula of the form
\begin{equation} \label{053a}
{\partial\greenregperp_\smooth^\dimqmp\over \partial r^a}\{r^2\}
=- r_a\, {\calJ}_\smooth^\dimqmp\{r^2\}\, ,
\end{equation}
in which the function ${\calJ}_\smooth^\dimqmp\{r^2\}$ is defined as 
the integral
\begin{equation}\label{054}
{\calJ}_\smooth^\dimqmp\{r^2\} =
\Opm1\int_0^\infty\!\!\sigma^{p-1} {{\calI}_\smooth^\dimq
\{r^2\!+\sigma^2\}\over (r^2\!+\sigma^2)^{q/2}}\, \dd\sigma \,,
\end{equation}
which will always converge by virtue of (\ref{018}). 

In a similar manner, starting from (\ref{013}), it can  be seen that 
the corresponding second derivative components will be given in terms of 
the same convergent integral ${\calJ}_\smooth^\dimqmp\{r^2\}$ by
\begin{equation}\label{055}
{\partial^2\widetilde\phi_\smooth\over\partial r^a\partial
 r^b}\{r^2\}=\bar\rho\, {\partial^2\greenregperp_\smooth^\dimqmp\over
 \partial r^a \partial r^b}\{r^2\}
\end{equation}
in which, by the definition (\ref{044}), it can be seen that we
shall have
\begin{equation}\label{055a}
{\partial^2\greenregperp_\smooth^\dimqmp\over \partial r^a
\partial r_b}\{r^2\}= 2 {\Opm1}
\int_0^{\cutoff} \sigma^{p-1}\left(\! \perp_{ab} 
{ \dd \greenreg_\smooth^\dimq
\over \dd s^2} +2 r_a r_b  { \dd^2 \greenreg_\smooth^\dimq
\over (\dd s^2)^2}\right)\{\sigma^2+r^2\} 
\, \dd\sigma \, ,
\end{equation}
in which it is to be recalled that that the use of curly brackets 
(as in the expression $\{\sigma^2+r^2\}$ at the end) indicates 
functional dependence (not simple multiplication). Using 
(\ref{016a}) to express the second derivative in the integrand as
\begin{equation} \label{055b}
{\dd^2\greenreg_\smooth^\dimq\over (\dd s^2)^2}=
{(p-q)\over 2 r^2}{\dd \greenreg_\smooth^\dimq
\over \dd s^2}- {\Oqm1\over 4r^2}\profil_\smooth^\dimq\{s^2\}
-{\sigma^{2-p}\over r^2} {\dd\over\dd\sigma^2}\left(\sigma^p
{\dd \greenreg_\smooth^\dimq \over \dd s^2}\right)\, .
\end{equation}
and observing that the asymptotic behaviour (\ref{019}) 
ensures that the last term in (\ref{055b}) will contribute nothing
to the integral on the right of (\ref{055a}), we finally obtain
an expression of the convenient form
\begin{equation}\label{055c}
{\partial^2\greenregperp_\smooth^\dimqmp\over \partial r^a
\partial r_b}\{r^2\} =-{r_a r_b\over r^2}\,\Oqm1
\profilperp_\smooth^\dimqmp\{r^2\} +\left[(q-p){r_a r_b\over r^2}
-\!\perp_{ab}\right] {\calJ}_\smooth^\dimqmp\{r^2\}\, ,
\end{equation}
in which $\profilperp_\smooth^\dimqmp\{r^2\}$ is the dimensionally
reduced profile function  given by the integral (\ref{042}).

\section{Illustration using a canonical profile function}\label{sec_cano}

Our main results will be derived for any generic smoothing ansatz,
without restriction to any particular form for the profile function
$\profil^\dimq_\smooth\{s^2\}$. However as a concrete illustration, 
it is instructive to consider the case of what we shall refer to as the 
``canonical'' regularisation ansatz whereby the $q$-dimensional Dirac 
delta ``function'' $\delta^\dimq\{x,\bar x\}$ and the associated Green 
function $\greenf^\dimq(x,\bar x)$ are considered as the respective 
limits -- as the regularisation parameter $\thickness\rightarrow 0$ --
of one-parameter families of smooth bi-scalar functions
\begin{equation}\label{061}
\profil_\smooth^\dimq\{x,\bar x\} =\delta^\dimq_\smooth\{s^2\}\,,
\qquad
\hbox{and}\qquad
\greenreg_\smooth^\dimq\{x,\bar x\} =\greenf^\dimq_\smooth\{s^2\}
\end{equation}
that are characterised in terms of the squared distance
$s^2\{x,\bar x\}$ and of the smoothing lengthscale $\thickness$ by the
prescription
\begin{equation}\label{005}
\delta^\dimq_\smooth\{s^2\}={\thickness^2 q \over\Oqm1}\left(s^2
+\thickness^2\right)^{-(q+2)/2} \, .
\end{equation}

This canonical profile function evidently satisfies
$\delta^\dimq_\smooth\{s^2\}\rightarrow 0$ when 
$\thickness^2\rightarrow\infty$ for $s^2\ne 0$, as well as the unit
normalisation condition (\ref{05}) using the fact that 
$\dd{\cal S}^{[q]}=\Oqm1 s^{q-1}\dd s$. By expressing the canonical profile
function as
\begin{equation}
\delta^\dimq_\smooth\{s^2\}={2\over\Oqm1}s^{2-q}{\dd\over \dd s^2}
\left[\left({s^2\over s^2+\thickness^2}\right)^{q/2}\right]\,,
\end{equation}
it can be seen that the resulting integral (\ref{017}) will simply be 
given by
\begin{equation}\label{5}
{\calI}^\dimq_\smooth\{s^2\}= \left(\frac{s^2}{s^2+\thickness^2}
\right)^{q/2}\, .
\end{equation}
It is then straightforward, by integrating (\ref{016}), to show
that the corresponding regularised Green function, satisfying by the
Poisson equation (\ref{07}), will be given by the canonical Green
function $\greenf^\dimq_\smooth$ defined by
\begin{equation}\label{7}
\greenf^\dimq_\smooth\{s^2\}={1\over(q-2)}
\big( s^2+\thickness^2 \big)^{-(q-2)/2}  \,,
\end{equation}
as long as $q>2$. As explained above, it is determined up to a constant
that is fixed by the fall off requirement at infinity which is mandatory
for the convergence of the integrals considered below but that will not
be relevant for the gradients to be computed later on. In the marginally
convergent case where $q=2$, one finds that
\begin{equation}\label{7b}
\greenf^{^{[2]}}_\smooth\{s^2\}=-\frac{1}{2}\ln\left\{1+\frac{s^2}
{\thickness^2}\right\} \,.
\end{equation}

For a specific physical system this particular choice is not necessarily 
the ansatz that will be most accurate for the purpose of providing a 
realistic representation of its actual internal microstructure. However 
this ``canonical'' prescription -- as characterised by the formulae
(\ref{5}) and (\ref{7}) -- is distinguished from other conceivable
alternatives by the very convenient property of having an analytic
form that is preserved by the dimensional reduction procedure.
The associated radial profile function is simply given by
\begin{equation}\label{07b}
\profilperp_\smooth^\dimqmp\{r^2\}=
\frac{\Opm1}{\Oqm1}\frac{\thickness^2 q}{(\thickness^2+r^2)^{(q-p+2)/2}}
\int_0^\infty\frac{u^{p-1}}{(1+u^2)^{(q+2)/2}} \dd u\,.
\end{equation}
Since we shall always have $q> p-2$, the latter integral will always 
be convergent. Using the well known properties (see (\ref{A1}) in the 
appendix) of the Euler Beta function $B\{p/2,(q-p)/2+1\}$ 
it can be seen that in this canonical case
the radial profile function will be given simply by
\begin{equation}\label{063a}
\profilperp^\dimqmp_\smooth\{r^2\}=\delta^{^{[q-p]}}_\smooth\{r^2\}
\, .
\end{equation}
It can similarly be seen that, since we always have $q>p$, the integral 
(\ref{054}) will also be convergent, and (again using the formulae in 
the appendix) that the result will be expressible in terms of the 
Beta function as
\begin{equation}\label{068}
{\calJ}_\smooth^\dimqmp\{r^2\}={\Oqmp\over
2(r^2\!+\thickness^2)^{(q-p)/2}} B\left\{\frac{p}{2},
\frac{q-p}{2}\right\}\, .
\end{equation}

According to (\ref{044}) the corresponding dimensionally reduced Green 
functions will have a canonical form
\begin{equation}\label{066a}
\greenregperp_\smooth^\dimqmp\{r^2\}=\greenfperp_\smooth^\dimqmp\{r^2\}
\end{equation}           
that will similarly be given whenever they are well defined
-- that is  whenever  $q>p+2$ -- by expressions of the form  
\begin{equation}\label{066}
\greenfperp_\smooth^\dimqmp\{r^2\}={\Opm1\over 2(q-2)
(\thickness^2\!+r^2)^{(q-p)/2-1}} B\left\{\frac{p}{2},
\frac{q-p}{2}-1\right\}\, ,
\end{equation}
so that, in terms of the canonical Green function (\ref{7}), they take
the form
\begin{equation} \label{067}
\greenfperp_\smooth^\dimqmp\{r^2\}=
{\Oqm1\over\Oqmpm1}\greenf_\smooth^{^{[q-p]}}\{r^2\}\, .
\end{equation}
The form of the Green function, unlike that of the canonical profile
function, is thus not exactly preserved by the dimensional reduction
process but the only difference is an overall volume factor. The
preceding expression is only valid when $q>p+2$, but in the marginally 
convergent hyperstring case where $q=p+2$ one can use the expression 
(\ref{7b}) for the canonical Green function to obtain the asymptotic 
relation
\begin{equation} \label{067b}
\greenfperp_\smooth^{^{[q,q-2]}}\{r^2\}\sim
\frac{\Omega^{^{[q-3]}}}{(q-2)}\ln\left\{\frac{\cutoff}{
\sqrt{r^2+\thickness^2} }\right\}
\end{equation}
as $\cutoff\rightarrow\infty$.  It is to be noted that the
logarithmic term in this formula and in (\ref{052}) will be 
expressible by (\ref{7b}) in the form
\begin{equation} \label{067c}
\ln\left\{\frac{\cutoff}{\sqrt{r^2+\thickness^2} }\right\}
=\greenf_\smooth^{^{[2]}}\{r^2\}+\ln\left\{\frac{\cutoff}{\thickness}\right\}
\end{equation}
in which the first term is independent of the cutoff $\cutoff$.
This means that in the asymptotic limit the dependence on
$r$ will drop out, so that we shall be left simply with
\begin{equation} \label{067d}
\greenfperp_\smooth^{^{[q,q-2]}}\{r^2\}\sim
\frac{\Omega^{^{[q-3]}}}{(q-2)}\ln\left\{\frac{\cutoff}{
\thickness }\right\}
\end{equation}
as $\cutoff\rightarrow\infty$.

\section{Source weighted averages in flat canonical case}\label{sec_weight}

For the purposes of a macroscopic description, what really matters
is not the detailed field distribution in a smoothed microscopic
treatment as developed in the previous sections, but only its effective 
averaged values. For instance, in the example of the scalar field 
$\widetilde\phi_\smooth\{r\}$, we shall be interested in its 
cross-sectional average defined as
\begin{equation}\label{b1}
\left\langle\widetilde\phi_\smooth\right\rangle\equiv\Oqmpm1
\int_0^\infty r^{q-p-1}\widetilde\phi_\smooth\{r\} w\{r\}\, \dd r\,,
\end{equation}
where $w\{r\}$ is a weighting factor dependent only on the
radial coordinate $r$ subject to  the normalization condition
\begin{equation}
\Oqmpm1\int_0^\infty r^{q-p-1}w\{r\}\,\dd r =1 \,.
\end{equation}

Let us first consider the flat uniform configurations studied in
Section~\ref{sec_flat}. Whatever the explicit form of the function 
$w\{r\}$, it is obvious that the isotropy of the configuration will
ensure that the average of the field gradient (\ref{053}) will cancel 
out, meaning that we shall have $\left\langle\partial
\widetilde\phi_\smooth/\partial r^a\right\rangle=0$.
Using the general relation 
\begin{equation}
\left\langle\frac{r_ar_b}{f\{r\}}\right\rangle
=\frac{1}{q-p}\perp_{ab}\left\langle\frac{r^2}{f\{r\}}\right\rangle\,,
\end{equation}
which applies to  any purely radial function $f\{r\}$, it can thus be 
seen from (\ref{055}) and (\ref{055c}), using the expression
(\ref{041}) for the radial source density distribution, that 
\begin{equation}\label{b2}
\left\langle\frac{\partial^2\widetilde\phi_\smooth}
{\partial r^a\partial r^b}\right\rangle=- \frac{\Oqm1}{q-p}
\left\langle\widetilde\rho_\smooth\right\rangle\perp_{ab}\,,
\end{equation}
in which the value of $\left\langle\widetilde\rho_\smooth
\right\rangle$ also depends on the choice of the weighting function. 
It is to be remarked that (as a useful check on the preceding algebra)
the trace this equation gives back the original Poisson equation  
(\ref{055a}).

The kinds of energy and force contributions that are relevant in 
physical applications will typically be proportional to the product 
of the linear field $\phi$ with the corresponding source term. 
This implies that the weighting of the field $\widetilde\phi_\smooth$ 
considered here should be proportional to the source term
$\widetilde\rho_\smooth$. For a flat configuration, it can thus be
deduced from (\ref{041}) that the appropriate weighting factor
should be explicitly given by the formula
\begin{equation}\label{58}
w\{r\}=\profilperp^\dimqmp_\smooth\{r^2\}\,,
\end{equation}
and hence that the average of the source term will be
given by an expression of the form
\begin{equation}\label{a1}
\left\langle\widetilde\rho_\smooth\right\rangle=\bar\rho\Oqmpm1
\int_0^\infty  \left[\profilperp^\dimqmp_\smooth\{r^2\}\right]^2 
r^{q-p-1}\,\dd r\,,
\end{equation}
which will be valid, whatever the profile function, as long as the 
configuration under consideration is flat. The corresponding
expression for the average of the field $\phi$ will have the form
\begin{equation}\label{a2}
\left\langle\widetilde\phi_\smooth\right\rangle=\bar\rho\Oqmpm1
\int_0^\infty \profilperp^\dimqmp_\smooth\{r^2\}
\greenregperp^\dimqmp_\smooth\{r^2\} r^{q-p-1}\, \dd
r\,.
\end{equation}

In the special canonical case introduced in Section~\ref{sec_cano}, 
the profile factor will be given by (\ref{005}), so that the (always 
convergent) integral (\ref{a1} will be obtainable explicitly --
as shown in the appendix -- in terms of the Beta function.
This leads to an expression of the form
\begin{equation}
\left\langle\rho_\smooth\right\rangle=\frac{(q-p)^2}{2\Oqmpm1}
B\left\{\frac{q-p}{2},\frac{q-p}{2}
+2\right\}\frac{\bar\rho}{\thickness^{q-p}}\,
\end{equation}
for the average source density. With further use of the formulae 
in the appendix, this can be written even more explicitly as
\begin{equation}\label{59}
\left\langle\rho_\smooth\right\rangle=\frac{1}{16\pi^{q-p\over 2}}
\frac{(q-p)^2(q-p+2)}{(q-p+1)}\frac{[\Gamma\{(q-p)/2\}]^3}
{\Gamma\{q-p\}} \frac{\bar\rho}{\thickness^{q-p}} \,  .
\end{equation}
The simplest application of this result is to the case of a
hypermembrane or ``wall'', meaning a hypersurface forming brane as
characterised by $q-p=1$, for which the formula (\ref{59}) just gives
$\left\langle\rho_\smooth\right\rangle=3\pi\bar\rho/(32\thickness)$. 
For the case $q-p=2$, which includes that of a string in an ordinary
3-dimensional space background, the corresponding result is
$\left\langle\rho_\smooth\right\rangle=\bar\rho/(3\pi \thickness^2)$.

In terms of the canonically reduced Green functions (\ref{066a}), it
it can be seen that the average of the the regularised field
$\phi_\smooth$ will be given in the flat case by a prescription of 
the form
\begin{equation}
\left\langle\phi_\smooth\right\rangle=\bar\rho\Oqmpm1\int_0^\infty
\greenfperp_\smooth^{\dimqmp}\{r^2\}\delta_\smooth^{^{[q-p]}}\{r^2\}
r^{q-p-1}\dd r\,.
\end{equation}

Whenever the expression (\ref{067}) is valid, that is whenever $q>p+2$, 
the formulae in the appendix will provide a  result expressible as
\begin{equation}
\left\langle\phi_\smooth\right\rangle=\frac{1}{2}
\frac{(q-p)}{(q-p-2)}\frac{\Oqm1}{\Oqmpm1}B\left\{\frac{q-p}{2},
\frac{q-p}{2}\right\}\frac{\bar\rho}{\thickness^{q-p-2}}\,,
\end{equation}
which can be rewritten in terms of $\Gamma$ functions as
\begin{equation}\label{79}
\left\langle\phi_\smooth\right\rangle=\frac{\pi^{p/2}}{2}
\frac{(q-p)}{(q-p-2)}
\frac{\left[\Gamma\{(q-p)/2\}\right]^3}{\Gamma\{q/2\}
\Gamma\{q-p\} } \frac{\bar\rho}{\thickness^{q-p-2}}\,.
\end{equation}

The preceding formula will no longer be valid, but an analogous though  
weakly cutoff dependent expression will still be obtainable in in the 
marginal hyperstring case, when $q=p+2$. In this case, instead of 
(\ref{067}), the effective dimensionally reduced Green function will be 
given by the simple asymptotic formula (\ref{067d}) in the large value 
limit for the relevant infrared cutoff $\cutoff$. The ensuing asymptotic 
formula for the averaged value of $\phi_\smooth(r)$ in a
hyperstring is thus immediately seen to be given by
\begin{equation}
\left\langle\phi_\smooth\right\rangle\sim
\frac{{\bar\rho}\,\Omega^{^{[q-3]}}}{2(q-2)}\ln\left\{\frac{\cutoff}
{\thickness}\right\}
\end{equation}
as $\cutoff\rightarrow\infty$.

\section{Allowance for curvature in the averaging process}
\label{sec_curv_avrg}

We are now ready to tackle the main problem that motivates this study, 
which is the evaluation of the effect of curvature of the brane 
$p$-surface.

In order to do this, instead of the exactly flat configuration
characterised by the coordinate system (\ref{31}) and the particular
form of the metric components (\ref{32}), we need to consider a
generically curved $p$-surface adjusted so as to be tangent to the
reference flat $p$-surface at the coordinate origin.

The validity of the ``thin brane'' limit description with which we 
are concerned depends on the requirement that the relevant macroscopic 
lengthscale $\cutoff$ (that was used to specify an appropriate 
``infra-red'' cut off, in cases for which the convergence condition 
$q>p+2$ does not hold) should be large compared with the  microscopic 
brane thickness scale $\thickness$. 
Subject to the  requirement that
\begin{equation} \label{60}
\petit\equiv{\thickness\over\cutoff}\ll 1  
\end{equation}
is satisfied (either because $\cutoff$ is very large or because
$\thickness$ is very small), it will be sufficient for our present
purpose to work at linear order in $\petit$. This means that we need 
only consider up to quadratic order in $\sigma^i$ for an expansion where 
the  brane locus will be specified in the relevant neighbourhood -- of
dimension large compared with $\thickness$ but small compared with
$\cutoff$ -- by
\begin{equation}\label{61}
\bar z^i =\sigma^i\, ,\qquad
\bar r^a= {1\over2}{K_{ij}}^a
\sigma^i\sigma^j\left[1+o\{\petit\right)\}]\, ,
\end{equation}
for an arbitrary set of constant coefficients ${K_{ij}}^a$.  In this
expansion, we have used the standard meaning of $o\{\delta\}$ denoting 
a quantity negligible with respect to $\delta$, that is satisfying
$o\{\delta\}/\delta\rightarrow 0$ in the limit $\delta\rightarrow 0$.  
It can be easily verified that, as our notation (\ref{61}) suggests, 
the constants of the development are exactly given by the non
vanishing components of the second fundamental tensor
${K_{\mu\nu}}^\rho$, defined in (\ref{16}) at the origin, with
respect to the orthonormal background coordinates (\ref{31}).

In the simplest cases, the condition (\ref{60}) might be satisfied
simply by taking $\cutoff$ to be arbitrarily large. In specific
applications an upper bound on the admissible magnitude for $\cutoff$
will commonly be imposed by the physical nature of the environment in
which upper limits might be provided by lengthscales such as those
characterising the gradients of relevant background fields or those
characterising the separation between neighbouring branes. 
However, an upper limit that must always be respected will
be provided, except in the strictly flat case, by the curvature
lengthscale defined by the inverse of the magnitude of the mean
curvature vector whose components are given by the formula
\begin{equation}\label{62}
K^a= {K_i}^{ia}\,.
\end{equation}
As our notation suggests, these components $K^a$ are identifiable
with respect to the orthonormal background coordinates (\ref{31})
as the  non-vanishing components of the complete curvature vector 
$K^\mu$ given by (\ref{19}). In order to ensure that none of the  
coefficients ${K_{ij}}^a$ exceeds the order of magnitude of the 
inverse lengthscale $\cutoff^{-1}$, the latter must thus be 
subject to the limitation
\begin{equation}\label{62a}
\cutoff^{-2}\gta K^a K_a\, .
\end{equation}
In particular, our formalism will break down near a kink or a cusp of 
the brane worldsheet.

The curvature of the worldsheet makes the evaluation of the physically
appropriate weighting factor for the definition of effective sectional
averages across the brane rather more delicate than in the flat case
dealt with in the preceding section. The kind of average that is
ultimately relevant is a source density weighted mean taken not just
over a $(q-p)$-dimensional orthogonal section through the brane but
rather over the $q$-dimensional volume between neighbouring
$(q-p)$-dimensional orthogonal sections through the brane. This
distinction is irrelevant for a flat brane for which neighbouring
sections are exactly parallel. However for a slightly curved brane,
there will be a relative tilt between nearby orthogonal cross sections,
which means that the effective weighting factor $w\{r\}$
appropriate for the purpose of averaging over a $(q-p)$-dimensional
cross-section will no longer be exactly proportional to the density
$\rho\{r\}$ as in the flat case. This needs to be adjusted by a correction
factor allowing for the relative compression or expansion of the
relevant $q$-volume due to the relative tilting. By summing over the
different directions in which the tilting can occur, it can be seen that
the weighting factor will be given, at first order,
by
\begin{equation}\label{63}
w\{r\}={\widetilde\rho_\smooth\!\{r\}\over\bar\rho}\left[1-K_a r^a +
o\{\petit\}\right] \, .
\end{equation}

To obtain the corresponding generalisation of the formula (\ref{58}) for
the flat case, we must allow for the fact that the original formula
(\ref{041}) for the source density distribution also needs to be
corrected to take into account the effect of the curvature.  Using
(\ref{61}) to expand the distance function (\ref{6}) as
\begin{equation}\label{64} 
s^2\{x,\sigma\}= r^2 +\sigma^2 -\perp_{ab}{K_{ij}}^{a}r^b
\sigma^i\sigma^j\left[1+ o\{\petit\}\right]\, ,
\end{equation}
we obtain the Taylor expansion
\begin{equation}\label{065}
\profil_\smooth^\dimq\{s^2\}=\profil_\smooth^\dimq\{r^2+\sigma^2\}-
{\dd\profil_\smooth^\dimq\{r^2+\sigma^2\} \over \dd\sigma^2}
\perp_{ab}{K_{ij}}^{a}r^b\sigma^i\sigma^j\left[1+o\{\petit\}\right]\, ,
\end{equation}
for the profile function (\ref{06}). Since we are neglecting
corrections of quadratic and higher order in the curvature, it can be
seen that when the integration is carried out in two stages of which the
first is just to take the integration over the $p$-sphere of constant
radius $\sigma$ centered on the origin, it is possible to
replace the surface element $ \dd{\cal S}^\dimp$ in (\ref{04}) by
$\Opm1\,\sigma^{p-1} \, \dd\sigma $ as in (\ref{042}).
Furthermore, in the anisotropic contribution proportional to
$\sigma^i\sigma^j$ it will be possible to replace $\sigma^i\sigma^j
\dd{\cal S}^\dimp$ by the spherically averaged 
equivalent $\gamma^{ij} \Opm1\,\sigma^{p+1}\, \dd\sigma/p $. Therefore,
to allow for the effect of the curvature at first order,
the density distribution (\ref{041}) needs to be replaced by
\begin{equation}\label{065a}
\widetilde\rho_\smooth\{r\}=\bar\rho\,\left[
\profilperp_\smooth^\dimqmp\{r^2\} 
-\frac{\Opm1}{p}\big[K_a r^a+o\{\petit\}\big] 
\int_{0}^{\infty}\sigma^{p+1}\dd\sigma \,{\dd
\profil_\smooth^\dimq\{r^2+\sigma^2\}\over\dd\sigma^2}\right] \, ,
\end{equation}
in which the second term can be evaluated using an integration by 
parts, using the definition~(\ref{042}) for 
$\profilperp_\smooth^\dimqmp\{r^2\}$ and the final result takes the 
simple form
\begin{equation}\label{68}
\widetilde\rho_\smooth\{r\}=\bar\rho\profilperp_\smooth^\dimqmp\{r^2\}
\left[1+ {1\over2} K_a r^a+o\{\petit\}\right]\, .
\end{equation}
Thus, it can be seen that the geometric curvature adjustment
factor in (\ref{63}) is partially canceled by the density adjustment
factor in (\ref{68}) to give the corresponding curvature adjusted 
weighting factor
\begin{equation}\label{69}
w\{r\}=\profilperp_\smooth^\dimqmp\{r^2\}
\left[1-{1\over2} K_a r^a+o\{\petit\}\right]\, .
\end{equation}

We note that  as a corollary of this partial cancellation, the first 
order effect of the curvature is entirely canceled in the 
correspondingly weighted density function, which will be given 
simply by 
\begin{equation}\label{69b}
w\{r\}\,\widetilde\rho_\smooth\{r\}= \bar\rho\left[
\profilperp_\smooth^\dimqmp\{r^2\}\right]^2
\left[1+o\{\petit\}\right]\, .
\end{equation}
Thus to first order in the curvature, the average density will 
still be given by the same form (\ref{a1}) as in the flat case.

\section{Allowance for curvature in the potential
gradient}\label{sec_potcurv}

In the flat configuration, the average of the field gradient
$\left\langle\partial\widetilde\phi_\smooth/\partial r^a\right\rangle$, vanishes.
Thus, unlike $\left\langle\widetilde\phi_\smooth\right\rangle$ and
$\left\langle\partial^2\widetilde\phi_\smooth/\partial r^a\partial
r^b\right\rangle$ for which the dominant contributions, respectively given
by Eqs.~(\ref{b1}) and (\ref{b2}), are of zeroth order in the
curvature, it will have a dominant contribution that is of linear
order in the curvature.

Starting from the general expression (\ref{09}) for the field
$\widetilde\phi_\smooth$ in terms of the generalised Green function, the
strict analogue of the integral (\ref{065}) for the source density $
\widetilde\rho_\smooth$ is given, after Taylor expanding the Green
function, by
\begin{equation}\label{071}
\widetilde\phi_\smooth\{r\}=\bar\rho\,\Opm1
\int_0^\cutoff\!\sigma^{p-1} \dd\sigma\left[\greenreg_\smooth^\dimq
\{r^2+\sigma^2\}-\frac{\sigma^2}{p} {\dd\greenreg_\smooth^\dimq
\{r^2+\sigma^2\}\over\dd\sigma^2}K_a r^a \right]
\left[1+o\{\petit\}\right]\,.
\end{equation}
The truncation of the integration at the finite infra-red cut-off
$\cutoff$ is only needed for the treatment of the marginally divergent
case for which $q=p+2$, but makes no difference, at first order in
$\thickness/\cutoff$. In the convergent cases $q>p+2$, the result will
be effectively the same as would be obtained simply by taking the
limit $\cutoff\rightarrow\infty$. The corresponding expression for the
gradient is found, after an integration by parts, to be
\begin{equation}\label{075}
{\partial\widetilde\phi_\smooth\over\partial r^a }= 
\bar\rho\,\Opm1\!\int_0^\cutoff\!\sigma^{p-1} \dd\sigma\left[
\left(2+K_b r^b\right)r_a - {\sigma^2\over p}K_a\right]
{\dd\greenreg_\smooth^\dimq\{r^2+\sigma^2\}\over \dd\sigma^2}
\left[1+o\{\petit\}\right]\,.
\end{equation}

The appropriately weighted potential function required for the
evaluation of the relevant average can now be seen from (\ref{69}) and
(\ref{071}) to be given, after another integration by parts, by
\begin{eqnarray}\label{82a}
w\{r\}\,\widetilde\phi_\smooth\{r\}=\bar\rho\,\Oqm1
\profilperp_\smooth^\dimqmp\{r^2\}\! 
\int_{0}^{\cutoff}\!\sigma^{p-1} \dd\sigma\, 
\greenreg_\smooth^\dimq\{r^2+\sigma^2\}\left[1+o\{\petit\}\right]\cr
-\bar\rho\,\Oqm1\profilperp_\smooth^\dimqmp\{r^2\} 
K_br^b {\cutoff^p\over 2p} \greenreg_\smooth^\dimq\{r^2\!+\cutoff^2\}
\, .
\end{eqnarray}
When $q>p+2$, the asymptotic behaviour (\ref{020}) implies that the
boundary contribution tends to zero, while the integral converges toward 
the radial Green function, as defined in (\ref{044}) so that we obtain 
an expression of the simple cut-off independent form
\begin{equation}
w\{r\}\,\widetilde\phi_\smooth\{r\}=\bar\rho\,
\profilperp_\smooth^\dimqmp\{r^2\}
\greenregperp_\smooth^\dimqmp\{r^2\}\,.
\end{equation}

In the marginal case $q=p+2$, we obtain an expression of the asymptotic
form
\begin{equation}
w\{r\}\,\widetilde\phi_\smooth\{r\}=\bar\rho\,
\profilperp_\smooth^{^{[q,q-2]}}\{r^2\}\left[\frac{\Omega^{^{[q-3]}}}
{q-2}\ln\left\{\frac{\cutoff}{\thickness}\right\}-\frac{K_ar^a}
{2(q-2)^2} \right]\,.
\end{equation}
in which the boundary contribution at the end will provide a finite 
curvature adjustment term. However, since this adjustment term is an 
odd function of the radius vector $r^a$ it will still provide no net 
contribution to the integrated average, which will thus be given 
simply by
\begin{equation}\label{82}
\left\langle\widetilde\phi_\smooth\right\rangle=\bar\rho\, \Oqmpm1
\!\int_0^\infty\! r^{q-p-1} \dd r \profilperp_\smooth^\dimqmp\{r^2\} 
\greenregperp_\smooth^\dimqmp\{r^2\}
\left[1+o\{\petit\}\right] \, ,
\end{equation}
which to first order in $\petit$ is just the same as in the 
corresponding formula (\ref{a2}) for the flat case.

When we go on to work out the analogously weighted potential gradient
distribution, it can be seen that the term with quadratic radial 
dependence in (\ref{075}) will cancel out, leaving
\begin{equation}\label{83}
w\{r\}{\partial\widetilde\phi_\smooth\over \partial r^a } = \bar\rho\,
\Opm1 \profilperp_\smooth^\dimqmp\{r^2\}\!
\int_{0}^{\cutoff}\!\sigma^{p-1} \dd\sigma \left[2r_a 
- {\sigma^2\over p}K_a\right]\! {\dd\greenreg_\smooth^\dimq
\{r^2+\sigma^2\}\over \dd\sigma^2}\left[1+o\{\petit\}\right]\, .
\end{equation}
Since the first term in the previous integral is an odd function of
the radius vector $r^a$, it provides no net contribution to the
corresponding integrated average, which will therefore be given simply
by
\begin{equation}\label{85}
\left\langle{\partial\widetilde\phi_\smooth\over\partial r^a}
\right\rangle= \bar\rho\,\Oqmpm1 K_a\int_0^\infty r^{q-p-1} \dd r 
\profilperp_\smooth^\dimqmp\{r^2\}\,{\calH}^\dimqmp\{r^2\}
\left[1+o\{\petit\}\right]\, ,
\end{equation}
where
\begin{equation}\label{86}
{\calH}^\dimqmp\{r^2\}\equiv-{\Opm1\over p}\int_{0}^{\cutoff}
\sigma^{p+1}\dd\sigma {\dd\greenreg_\smooth^\dimq\{r^2+\sigma^2\}
\over\dd\sigma^2} \, .
\end{equation}
Using an integration by parts, this integral can be rewritten in 
terms of the dimensionally reduced Green function
$\greenregperp_\smooth\dimqmp\{r^2\}$ defined by (\ref{044}) in the 
form
\begin{equation}\label{87}
{\calH}^\dimqmp\{r^2\}={1\over2}\greenregperp_\smooth^\dimqmp\{r^2\}
-{\Opm1\cutoff^p\over 2p}\,\greenreg_\smooth^\dimq\{r^2+\cutoff^2\}\, .
\end{equation}
As in (\ref{82a}), the boundary contribution at the end will vanish 
in the large $\cutoff$ limit in consequence of the asymptotic behaviour 
(\ref{020})  provided the convergence condition $q>p+2$ is satisfied. 
In the marginally divergent hyperstring case $q=p+2$, the boundary 
term will tend to a limit having a finite value (namely $-1/2p^2$) that 
will still be negligible compared with the first term in (\ref{87}). Thus, 
whenever $q-p\geq 2$ we shall always have a relation of the form 
\begin{equation}\label{88}
{\calH}^\dimqmp\{r^2\}={1\over2}\greenregperp_\smooth^\dimqmp
\{r^2\}\left[1+o\{\petit\}\right]\, .
\end{equation}

The only case to which this formula does not apply is that of a 
hypermembrane, meaning the hypersurface supported case characterised by 
$q=p-1$, for which the boundary term will be linearly divergent, like 
the corresponding the integral for $\profil_\smooth^\dimqmp\{r^2\}$
as characterised by (\ref{052a}). We can still derive an 
asymptotic relation
\begin{equation}\label{89}
{\calH}^{^{[q,q-1]}}\{r^2\}\sim {q-2\over 2(q-1)}
\greenregperp_\smooth^{^{[q,q-1]}}\{r^2\}
\end{equation}
as $\cutoff\rightarrow\infty$ even in this extreme case,
but its utility is limited by the strongly cut off dependence
of the quantities involved. It can be seen that it approaches
agreement with the generic formula (\ref{88}) when the
space dimension $q$ is very large. 

When (\ref{88}) is substituted back into (\ref{85}) one
obtains a result that can immediately be seen by comparison with
the formula (\ref{82}) for the averaged potential
$\left\langle\widetilde\phi_\smooth\right\rangle$ to be
expressible in terms of the latter by
\begin{equation}\label{90}
\left\langle{\partial\widetilde\phi_\smooth\over\partial r^a}
\right\rangle= 
{1\over2}K_a\left\langle\widetilde\phi_\smooth\right\rangle
\left[1+o\{\petit\} \right]\, ,
\end{equation}
a simple and easily memorable result whose derivation is the
main purpose of this work.

\section{Discussion}\label{sec_concl}

The preceding work shows not only that the relation (\ref{90}) holds 
as an ordinary numerical equality for the strictly convergent cases 
for which the co-dimension, $(q-p)$, is greater than 2, but also that 
it holds as an asymptotic relation for the marginally divergent case
of a hyperstring with $q-p=2$, including the previously studied examples
\cite{carter97,carter98,carter00} of application to a string with $p=1$ 
in the ordinary case of a background with space dimension $q=3$.

It is also clear from the preceding work that the formula (\ref{90})
is not applicable to the case of hypermembrane, for which the
co-dimension is given by $q-p=1$. In this case, it can be seen from 
(\ref{89}) that the factor $1/2$ in (\ref{90}) should in principle be 
replaced by a factor $(p-1)/2p$. However it is debatable whether 
any useful information can be extracted in this case since the result 
is strongly dependent on the cut-off due to the ``infra-red'' divergence,
and in practice this issue is of little importance because the
particularly good ultraviolet behaviour of the hypermembrane case
makes it easily amenable to other, more traditional, methods such as
the use~\cite{Battye01} of Israel Darmois type jump conditions.

Dropping the explicit reminder that higher order adjustments would
be needed if one wanted accuracy beyond first order in the ratio
of brane thickness to curvature radius, the relation (\ref{90})
translates into fully covariant notation as
\begin{equation}\label{81}
\left\langle\perp^{\mu\nu}\nabla_{\nu}\widetilde\phi_\smooth
\right\rangle={1\over2}K^\mu\left\langle\widetilde\phi_\smooth
\right\rangle\, .
\end{equation}
It is to be emphasised that the generic relation (\ref{81}) is a
robust result, whose validity, whenever the co-dimension is 2 or more,
does not depend on the particular choice of the canonical
regularisation ansatz on which the explicit formula (\ref{79}) for
$\left\langle\widetilde\phi_\smooth\right\rangle$ was based. It also 
shows that (\ref{81}) is valid for any alternative non-canonical
regularisation ansatz, whose mathematical properties would be less
convenient for the purpose of explicit evaluation, but that might
provide a more exact representation of the internal structure for
particular physical applications in cases where information about this
actual internal structure is available.

The question that remains to be discussed is how this result
generalises from static Poisson configurations, as considered here, to
dynamic configurations in a Lorentz covariant treatment for which the
Laplacian operator would be replaced by a Dalembertian operator. 

A priori, considerations just of Lorentz invariance imply that the
only admissible alternative to the formula (\ref{81}) would be a
formula differing just by replacement of the factor $1/2$ by some
other numerical pre-factor. However, it must be taken into account that
this numerical coefficient must match the coefficient that applies to
any static configuration. It can thus be deduced from the postulate of
consistency with what has been derived here that the formula
(\ref{81}) should be applicable to any $p$-brane with
$p+1$-dimensional timelike worldsheet in Lorentzian $q+1$-dimensional
background spacetime, for all values of $p$ and $q$ for which our
present derivation applies, that is to say for  $q-2\geq p\geq 1$.

There are two interesting opposite extreme cases that are beyond this 
range. At one extreme, in the most strongly ``infra red'' divergent 
hyper-membrane case $p=q-1$, the formula (\ref{81}) needs a modified 
factor $(p-1)/2p$. However, as discussed above, the quantities involved 
are, in this case, too highly cut off dependent for the result in question 
to have much significance, and anyway, as remarked above, their 
ultraviolet misbehaviour is so mild that hypermembranes can be 
satisfactorily treated without recourse to regularisation, so no 
analogue of the gradient formula is needed. 

On the other hand, at the opposite extreme, in the most strongly ``ultra 
violet'' divergent case, namely that of a simple point particle with 
$p=0$, an appropriate regularised gradient formula is indeed something
that is obviously  needed. However, for a static configuration, the 
possibility of curvature does not arise at all in this ``zero-brane'' case, 
so that -- while there is no reason to doubt the conjecture that it should 
still apply -- there is no short cut whereby the formula (\ref{81}) can be 
derived without further work, which will require recourse to the 
technical complications (due to the dimension sensitive nature of the time-dependent Green functions~\cite{CH}) involved in a fully dynamical treatment.

\section*{Appendix: Integral formulae and notation}
\label{apA}

The standard definition of the Euler Gamma function is provided
by the integral formula
\begin{equation}
\Gamma\{z\}\equiv\int_0^\infty \hbox{e}^{-t}t^{z-1}\dd t.
\end{equation}
When the argument has an integer value $n$ it can be shown,
using the obvious recursion relation $\Gamma\{n+1\}=n\Gamma\{n\}$, 
that it
satisfies the well known relations
\begin{equation}
\Gamma\{n+1\}=n!\, ,\qquad
\Gamma\left\{n+\frac{1}{2}\right\}=\frac{(2n)!}{2^{2n}n!}\sqrt{\pi}.
\end{equation}

In this article, we frequently need to evaluate integrals
of the form 
\begin{equation}
I_{p,q}\equiv\int_0^\infty\frac{y^{p-1}}{(1+y^2)^{q/2}}\dd y\,.
\end{equation}
Whenever $p>1$ and $q>p$, this integral will be convergent and
will be expressible  -- via a change of variables $t=1/(1+y^2)$ --
in the form
\begin{equation}
I_{p,q}=\frac{1}{2}B\left\{\frac{p}{2},\frac{q-p}{2}\right\}\,.
\end{equation}
where $B\{a,b\}$ is an Euler integral of the first kind
-- also known as the Beta function -- that is specified by
the formulae
\begin{equation}
B\{a,b\}\equiv\int_0^1 t^{a-1}(1-t)^{b-1}\dd t=
{\Gamma\{a\}\Gamma\{b\}\over\Gamma\{a+b\}}\,.
\end{equation}
The values, as given by (\ref{2}), of the surface area $\Oqm1$ of 
a unit $q$-sphere are related for different values of $q$ by
the formulae
\begin{equation}\frac{2\Oqm1}{\Opm1\Oqmpm1}
=\frac{\Gamma\left\{\frac{p}{2}\right\}
\Gamma\left\{\frac{q-p}{2}\right\}}{\Gamma\left\{\frac{q}{2}\right\}}
=B\left\{\frac{p}{2},\frac{q-p}{2}\right\}
\,,
\end{equation}
and thus satisfy the useful relation
\begin{equation}\label{A1}
\frac{q}{2}B\left\{\frac{p}{2},\frac{q-p}{2}+1\right\}
\frac{\Opm1}{\Oqm1}=\frac{q-p}{\Oqmpm1}.
\end{equation}
\bigskip

In order to facilitate the reading of the article, we conclude by
providing the following table summarising  the
the notation used  for the profile, Green,and other related
functions in the three cases considered in this article.\\

\begin{tabular}{|l|c|c|c|}
\hline
               & General notation   & Infinitely thin limit & canonical
               case\\
\hline

Potential      &  $\widetilde\phi_\smooth$      & $\phi$      &
               $\phi_\smooth$\\
Source term    &  $\widetilde\rho_\smooth$      & $\rho$      &
               $\rho_\smooth$\\
Profile distribution & $\profil_\smooth^\dimq$ & $\delta^\dimq$ &
               $\delta_\smooth^\dimq$\\
Green function  & $\calG_\smooth^\dimq$ & $\greenf^\dimq$ &
               $\greenf_\smooth^\dimq$\\
Radial profile function &  $\profilperp_\smooth^\dimqmp$ & 
$\delta^\dimqmp$ &               $\delta_\smooth^\dimqmp$\\
Radial Green function  & $\greenregperp_\smooth^\dimqmp$ & 
$\greenfperp^\dimqmp$ &     $\greenfperp_\smooth^\dimqmp$\\
\hline
\end{tabular}

\section*{Acknowledgments:} 
R.A.B. is funded by PPARC.



\begin{thebibliography}{000}

\bibitem{carter97}
B. Carter,
``Electromagnetic self-interaction in strings'',
{\it Phys. Lett.} {\bf B404} (1997) 246-252.
[hep-th/9704210].

\bibitem{carter98}
B. Carter and R.A. Battye,
``Non-divergence of gravitational self-interactions for Goto-Nambu strings'',
{\it Phys. Lett.} {\bf B430} (1998) 49-53.
[hep-th/9803012]. 

\bibitem{carter00}
B. Carter,
``Cancellation of linearised axion-dilaton self-interaction in strings'', 
{\it Int. J. Theor. Phys.} {\bf 38} (1999) 2779-2804.
[hep-th/0001136].

\bibitem{BW}
L. Randall and R. Sundrum,
``A Large Mass Hierarchy from a Small Extra Dimension'',
{\it Phys. Rev. Lett.} {\bf 83} (1999) 3370-3373.
 [hep-ph/9905221]

\bibitem{Davis01}
 A-C. Davis, I. Vernon, S.C. Davis, and W.B. Perkins,
``Brane World Cosmology Without the Z2 Symmetry'',
{\it Phys. Lett.} {\bf B504} (2001) 254-261.
[hep-ph/0008132]

\bibitem{CU01}
B. Carter and J-P. Uzan
``Reflection symmetry breaking scenarios with minimal gauge form coupling 
in brane world cosmology'',
 {\it Nucl. Phys.} {\bf B606} (2001) 45-58.
[gr-qc/0101010]

\bibitem{high}
A.G. Cohen and D.B. Kaplan, 
``Solving the Hierarchy Problem with Noncompact Extra Dimensions'',
{\it Phys. Lett.} {\bf B470} (1999) 52-58. 
[hep-th/9910132]

\bibitem{high2} 
R. Gregory, 
``Nonsingular global string compactifications'',
{\it Phys. Rev. Lett.} {\bf 84} (2000) 2564-2567. 
 [hep-th/9911015] 

\bibitem{high3}
T. Gherghetta and M. Shaposhnikov, 
 ``Localizing Gravity on a String-Like Defect in Six Dimensions'',
{\it Phys. Rev. Lett.} {\bf 85} (2000) 240-243. [hep-th/0004014]

\bibitem{carter95}
B. Carter,
``Dynamics of cosmic strings and other brane models'',
 in {\it Formation and Interactions of Topological Defects,
NATO ASI B349}, ed. A-C. Davis, R. Brandenberger
(Plenum, New York, 1995) 303-348. 
[hep-th/9609041].

\bibitem{Battye01}
R.A. Battye and B. Carter,
``Generic junction conditions in brane-world scenarios'',
{\it Phys. Lett.} {\bf B509} (2001) 331-336.
[hep-th/0101061]

\bibitem{CH}
R. Courant and D. Hilbert, {\it Methods of Mathematical Physics II : Partial Differential Eqautions}, Pg 681-698, (Wiley, New York, 1962).

\end{thebibliography}
\end{document}